\documentclass[prl,aps,twocolumn,superscriptaddress]{revtex4-1}
\usepackage{latexsym}
\usepackage{graphicx}
\usepackage{multirow}

\newcommand{\be}{\begin{equation}}
\newcommand{\ee}{\end{equation}}
\newcommand{\ba}{\begin{eqnarray}}
\newcommand{\ea}{\end{eqnarray}}
\newcommand{\baa}{\begin{eqnarray*}}
\newcommand{\eaa}{\end{eqnarray*}}

\def\be{\begin{equation}}
\def\ee{\end{equation}}
\def\bea{\begin{eqnarray}}
\def\eea{\end{eqnarray}}

\def\C60{A$_x$C$_{60}$}

\def\HgCu3{HgCa$_2$Cu$_3$O$_{8+y}$}
\def\HgCu4{HgBa$_2$Ca$_3$Cu$_4$O$_{10+y}$}
\def\TlCu{Tl$_2$Ba$_2$CuO$_{6+\delta}$}
\def\TlCu3{Tl$_2$Ba$_2$Ca$_2$Cu$_3$O$_{10+y}$}
\def\TlCu4{Tl$_2$Ba$_2$Ca$_3$Cu$_4$O$_{12+y}$}

\def\BiCu3{Bi$_2$Sr$_2$Ca$_{2}$Cu$_3$O$_y$}

\def\8LSCO{La$_{1.88}$Sr$_{.12}$CuO$_4$}
\def\110LNSCO{La$_{1.5}$Nd$_{0.4}$Sr$_{0.1}$CuO$_{4}$}
\def\stage4LCO{La$_{2}$CuO$_{4+\delta}$}
\def\Y248{YBa$_2$Cu$_4$O$_8$}

\def\NbSe2{NbSe$_2$}
\def\TaSe2{TaSe$_2$}
\def\TiSe2{TiSe$_2$}

\begin{document}

\title{Pressure Effects on Magnetically-Driven Electronic Nematic States in Iron-Pnictides }

\author{Jiangping Hu}
\affiliation{Beijing National
Laboratory for Condensed Matter Physics, Institute of Physics,
Chinese Academy of Sciences, Beijing 100080,
China}
\affiliation{Department of Physics, Purdue University, West
Lafayette, Indiana 47907, USA} 
\author{Chandan Setty}
\affiliation{Department of Physics, Purdue University, West
Lafayette, Indiana 47907, USA}
\author{Steve  Kivelson}
\affiliation{Department of Physics, Stanford University, Stanford, CA 94305}

\begin{abstract}
In a magnetically driven electronic nematic state,  an externally applied %nematic force created by 
uniaxial strain % can  %adequately  alter 
rounds the % (higher temperature)
 nematic transition and increases the magnetic transition temperature. %We derive such an effect
We study both effects in a simple classical model %for
of the iron-pnictides % based on an effective
expressed in terms of  local $SO(N)$ spins (with $N=3$) %magnetic exchange  model %within 
which we solve to leading order in $1/N$.
%using large N methods. 
The magnetic transition temperature is shown to increase linearly %with
in  response to an external %nematic force in the nematic state. A 
strain while a sharp crossover, %behavior of a 
which is a remnant of the nematic transition, can %still 
only be identified for extremely small strain.  %external force. 
  We show that these results 
  can reasonably account for %  explain  
  recent neutron experimental data in BaFe$_2$As$_2$ by C. Dhital {\it et al} \cite{dhital}.
\end{abstract}

\maketitle

% experimental nodal
 
%{\it Introduction:} %The u
Undoped %or 
and under-doped iron-pnictides  universally %possess
exhibit   colinear antiferromagnetic %ally (AF)  ordered %
 ground states %,  a collinear-AF
  (C-AF) %state  with  ordered
  with ordering
wavevectors %being either 
$(0,\pm\pi)$ or $(\pm\pi,0)$ with respect to the
tetragonal iron lattice.  The %development of the
 C-AF order is %always
necessarily accompanied %with 
by an orthorhombic %-tetragonal 
lattice distortion\cite{La1,Ce1}. 
The magnetic transition temperature $T_{AF}$ and the structural or ``nematic'' transition temperature $T_{\cal N}$ are closely related;  $T_{\cal N}$ is either equal to or slightly %above
greater than $T_{AF}$, %namely, 
$T_{\cal N}\geq T_{AF}$\cite{La1,Ce1}. 
The %structural 
nematic distortion breaks the $C_4$ rotational symmetry of the tetragonal lattice.\cite{comment}  %Meanwhile, 
The fact that the C-AF state  also breaks the same $C_4$
%rotational
 symmetry %. This correspondence 
 suggests the driving force of symmetry breaking may be %originated  electronically
 the magnetism, itself\cite{kivelson2008,xms2008,huxureview}, {\it i.e.} %. Namely, 
 % the symmetry breaking 
 the broken symmetry state % is
 should be thought of as an electronic nematic\cite{fradkin-nematic}. % state.
 %, a %prevailing common phenomenon in %many correlated electronic systems\cite{}. 

 %Recently, % this proposal has gained many experimental supports. 
 A number of striking experimental observations have been successfully interpreted in this light.
 %The t
 Transport measurements
reveal the existence of a large, intrinsic anisotropy  in the in-plane
resistivity above %the magnetic %Neel transition temperature
$T_{AF}$\cite{nematic-fisher2011, ying2011chen}. Similar
anisotropies of various physical quantities  % appear in %a variety of experimental
% measurements,
%including 
have been %seen in various physical quantities 
observed in
%local electronic structures measured by 
scanning
tunneling microscopy(STM)\cite{nematic-chuang2010}, magnetic
%fluctuations by 
neutron
scattering\cite{Zhaojun2009,Harringer2010}, %dynamic conductivity by 
optical reflectivity measurements \cite{Nakajima-nematic-optical}
and %electronic structures measured by 
angle resolved photoemission
spectroscopy (ARPES) \cite{nematic-yi2010,Yi2011-nematic} in
de-twinned 
%SAK samples
and (even at $T>T_{\cal N}$) strained samples.

However,  the origin of the nematic state %critical 
remains controversial. %  issues still remain regarding %the origin of the nematic state. 
As these materials are all metallic, some approaches emphasize the role of itinerant electrons, but the ``bad metal'' character of the conducting state, the small size of the Fermi pockets, and the relatively high scale of the ordering temperatures $T_{AF}$ and $T_{\cal N}$ suggest that a description in terms of localized classical spins (or possibly orbital moments), which neglects the itinerant electrons, may be sufficient to capture the essential physics.  (As the ordering temperatures are tuned toward $T=0$, where quantum effects become increasingly important and RKKY-like induced interactions become increasingly long-ranged, it certainly becomes increasingly problematic to ignore the effects of itineracy.)  At a minimum, the  large size of the ordered moments (which can exceed 1$\mu_B$ at low $T$), and the persistently commensurate character of the ordering rules out a picture of the ordered state based on a weak-coupling description and  Fermi-surface nesting.  It is also open to debate the extent to which lattice effects ({\it e.g.} electron-phonon coupling) and orbital ordering are essential drivers of the physics.  
%The controversy arises because of the existence of the interplay between multi-degrees of freedom and the complexity of electronic structures in iron-pnictides. The lattice, spin, and orbital degrees of freedom all manifest themselves in the nematic state, which causes a debate on the origin of the rotational symmetry breaking of the tetragonal lattice. Besides the lattice distortion and magnetic transition which break the $C^4$, an onsite ferro-orbital order between $d_{xz}$ and $d_{yz}$ orbitals also breaks the same $C^4$ rotational symmetry. Recently, such
For instance, 
%ferro-orbital order 
a small but evident difference in occupancy of the $d_{xz}$ and $d_{yz}$ in the nematic state has been observed by
ARPES\cite{nematic-yi2010,Yi2011-nematic} in %pressure 
de-twinned
samples under strain. %From standard symmetry argument, all the orders that break the same symmetry are allowed to be coupled directly with each other. The existence of any one of them thus could lead to the presence of the others. Therefore, it is difficult to disentangle the role of the different degrees of freedom behind the rotational symmetry breaking.
However, it follows from symmetry that any correlation function which transforms like the nematic order parameter will develop a non-zero expectation value in the nematic state, whether it is essential to the mechanism, or simply responding parasitically to the broken symmetry.

Recent neutron %experimental
scattering  data %in 
from BaFe$_2$As$_2$ by C. Dhital et al\cite{dhital} show that the C-AF magnetic transition can be affected  %under 
by relatively small strain fields. In this paper, %we show that this intriguing result is a direct evidence of the existence of an magnetic nematic order and can be explained in the model previously proposed  based on  effective local magnetic exchange couplings.
%Our calculation shows that due to the presence of the magnetic nematic order,  an external nematic force created by uniaxial strain can cause an enhancement of $T_{AF}$. The increase of $T_{AF}$ is linear with response to an external nematic force. The nematic transition in the presence of external strain field can be still identiffied as  a crossover behavior when the field is small.  
we adopt the most economical model which possesses the requisite ordered phases consisting of classical, localized, $SO(N)$ spins (with $N=3$ corresponding to the physically relevant Heisenberg case) residing on the Fe lattice with appropriately chosen antiferromagnetic couplings.  We solve this problem to leading order in $1/N$ in the large $N$ limit, including the effects of a small, externally imposed uniaxial strain.  We find that in response to a small uniform strain of magnitude $A_0$, the magnetic ordering temperature shifts
%SAK
\cite{fernandez}   according to
\begin{eqnarray}
\Delta T_{AF} \sim \left\{
\begin{array}{cc}
|A_0|^{1/\gamma}  &  {\rm if} \ T_{\cal N} = T_{AF}  \\
\chi \ A_0  &    {\rm if} \ T_{\cal N} > T_{AF}  
\end{array}
\right .
\label{chi}
\end{eqnarray}
where the susceptibility exponent $\gamma=2+{\cal O}(1/N)$ , and $\chi \sim (T_{\cal N}-T_{AF})^{-\gamma}$ as $T_{\cal N}\to T_{AF}$.  So long as $T_{\cal N} > T_{AF}$, the rounding of the %structural 
nematic transition occurs on a scale
\begin{eqnarray}
\Delta T_{\cal N} \sim |A_0|^x
\end{eqnarray}
where, again for $N\to \infty$, $x=1+{\cal O}(1/N)$.
 That we obtain results that satisfactorily account for the observations of Dhital {\it et al} supports the notion that this minimal model captures much of the essential physics of magnetism and nematicity in these materials. 

{\it Model:} 
%We start with the  minimal effective Hamiltonian that has been proposed to describe the magnetism in iron-pncitides,  the 
We start with the  previously considered $J_1-J_2-J_z-K$ model of magnetism in iron-pncitides:
\begin{eqnarray*}
H &=& \sum_{n,\vec r,\delta_1}[J_1 \vec{S}_{\vec r,n}\cdot\vec{S}_{\vec r+\delta_1,n} - K(\vec{S}_{\vec r,n}\cdot\vec{S}_{\vec r+\delta_1,n})^2]\\
&& + J_2 \sum_{n,\vec r,\delta_2}\vec{S}_{\vec r,n}\cdot\vec{S}_{\vec r+\delta_2,n} + J_z \sum_{n,\vec r}\vec{S}_{\vec r,n}\cdot\vec{S}_{\vec r,n+1}
\end{eqnarray*}
%SAK
in which $S_{\vec r,n}$ is a spin $S$ operator on the site $\vec r$ in %SAKthe 
 plane $n$ and $\delta_1$ and $\delta_2$ are the first and second nearest neighbor lattice vectors in the $Fe$ plane. $J_1$ and $J_2$ are the in-plane nearest neighbor (NN)  and next nearest neighbor (NNN)  magnetic couplings   respectively, $J_z$ is the coupling between layers along $c$ axis and $K$ is the NN biquadratic coupling.%SAKs.  
The C-AF groundstate arises for $J_2$ sufficiently large compared to $J_1$. (This condition reduces to $J_2 > J_1/2$ in the limit $S\to \infty$.) The origin of $K$ term has been discussed in \cite{Wysocki2011,magnetic-hu2011,huxureview}.   

To understand the finite temperature properties of this model analytically, we take the continuum limit and derive an effective classical field theory %of above Hamiltonian as shown in\cite{}
which captures the low energy physics of the above Hamiltonian\cite{kivelson2008}
\begin{eqnarray*}
H &&= N \int d^2 %\vec
 r \left\{\sum_{n,\alpha}\frac{\rho}{2} \mid \nabla \vec{\phi}_n^{(\alpha)}%(\vec r)
\mid^2 - %\tilde J
\frac {\rho_z} 2 \sum_{n,(\alpha)}\vec{\phi}_n^{(\alpha)}%(\vec r).
\cdot\vec{\phi}_{n+1}^{(\alpha)}%(\vec r)
\right .
  \\
&& \left .- \frac {Ng} {2}\sum_n \left(\vec{\phi}_n^{(1)}%(\vec r)
 \cdot\vec{\phi}_n^{(2)}%(\vec r)
 \right)^2 - \frac{\eta}{2}\sum_n\left(\partial_x\vec{\phi}_n^{(1)}\right)%(\vec r) 
 \cdot\left( \partial_y \vec{\phi}_n^{(2)}\right)%(\vec r),
 \right\}
\end{eqnarray*} 
where $\vec{\phi}_n^{(\alpha)}$ is a real $N=3$ component vector  field of unit norm [$\vec \phi_n(\vec r) \cdot \vec \phi_n(\vec r) =1$] representing the local orientation of the staggered magnetization on plane $n$ and sublattice $\alpha=1,2$ as defined in ref.\cite{kivelson2008}. Here %we have redefined 
 the couplings %as 
 are related to those in the Hamiltonian according to $\rho \propto J_2, \eta \propto J_1$, $%J
 \rho_z \propto J_z$\ and $g \propto K +\Delta K$, where %$K_0\sim 0.13\frac{J_1^2}{8J_2}$ ( We take S=1 for convenience without losing generality).
 $\Delta K\sim 0.13 \ J_1^2/SJ_2[ 1 + {\cal O}(1/S)]$ is  generated by quantum fluctuations of the spin.
%In the presence of 
An external strain fields $A_n(\vec r)$ %, there is a direct coupling between %nematic magnetic order and the fields. This coupling can be written as
induces a coupling between the sublattices:
\begin{eqnarray}
H \to H -%H_c=2 \sum_{\vec r} 
N\int d^2%\vec 
r\sum_n\ A_n(\vec r) \vec{\phi}_n^{(1)}(\vec r).\vec{\phi}_n^{(2)}(\vec r)
\end{eqnarray}
%  At the finite temperature, the total partition function then is given as
As a final step, we decouple the quardic term % via a Hubbard-Stratonovich transformation 
so that
\begin{equation}
Z= \int D\lambda%^1 D\lambda^2 D\sigma 
D\phi%^{(1)} D\phi^{(2)} 
\exp\left[-N\int d^2 %\vec
{r} \sum_n L_n\right]
\end{equation}
where the Lagrangian for each plane is given by
\begin{eqnarray*}
%\frac{L_n}{N} 
L_n&=&  \sum_{\alpha} i\lambda_n^{(\alpha)}\left( \mid \vec{\phi}_n^{(\alpha)}\mid^2 - 1\right) %+ 2
 - \left[\sigma_n+ \frac{A%(\vec r)
 }{T} \right]\left(\vec{\phi}_n^{(1)}%(\vec r).
\cdot\vec{\phi}_n^{(2)}\right)%(\vec r)
 \\
&&+ \frac{\rho}{T} \sum_{\alpha} \mid \nabla \vec{\phi}_n^{(\alpha)}%(\vec r)
\mid^2 -%J_c/T 
\frac {\rho_z} T\sum_{\alpha} \vec{\phi}_n^{(\alpha)}%(\vec r).
\cdot\vec{\phi}_{n+1}^{(\alpha)}%(\vec r)
\\
&& - \frac{\eta}{2T}\left(\partial_x\vec{\phi}_n^{(1)}\right)%(\vec r)
\cdot \left(\partial_y \vec{\phi}_n^{(2)}\right)%(\vec r)
+ \frac{T}{2Ng} \sigma_n^2
\end{eqnarray*}
where %SAKthe 
$\lambda_n^{\alpha}(\vec r)$ are the Lagrange
%SAKian 
multiplier%SAKs 
fields which enforce the normalization of $\vec \phi$ and $\sigma_n(\vec r)$ is the Hubbard-Stratonovich fields. 
For this action, %The 
the nematic order is given as $\langle \sigma_n\rangle=\frac g{NT}\langle \vec{\phi}_n^{(1)}%(\vec r)
\cdot\vec{\phi}_n^{(2)}%(\vec r)
\rangle $ and the magnetic order parameter by $\langle \vec \phi_n^{(1)}\rangle = \left[{\rm sign}(\langle \sigma_n\rangle) \right]\langle \phi_n^{(2)}\rangle$. 

The layered nature of the materials is reflected in the fact that we will always assume $\rho \gg \rho_z$ (in units in which the spacing between planes is 1), and we will shortly take the fact that the C-AF phase is most stable for $J_2 \gg J_1$ to justify neglecting the effects of the gradient coupling between the two sublattices, %{\it i.e.}
 we will set $\eta = 0$.\cite{note2}  The coupling constant $g$ determines the extent of separation between $T_{\cal N}$ and $T_{AF}$, which empirically is small implying that $g$, too, can be considered to be small.  To make this problem tractable, we will treat $1/N=1/3$ as a small parameter, %and
 {\it i.e.} we will report explicit results in the limit $N\to\infty$.  Generally, $N=3$ is large enough that no qualitative errors, and only small quantitative errors ({\it i.e.} in values of the critical exponents) are expected. 

For a 
constant strain field $A(\vec r)=A$, in the % large 
$N\to \infty$ limit, the nematic order can be obtained by finding the saddle point of the above Lagrangian, % determined by 
resulting in the following self-consistent equations for $\lambda$ and $\sigma$:
\begin{eqnarray*}
\sigma &=& \frac{g}{(2 \pi )^3} \int_{-\Lambda}^{\Lambda} dk_x \int_{-\Lambda}^{\Lambda}dk_y \int_0^{2 \pi} dk_z G_{12}(k_x,k_y,k_z)\\
1 &=& \frac{T}{(2 \pi )^3} \int_{-\Lambda}^{\Lambda} dk_x \int_{-\Lambda}^{\Lambda}dk_y \int_0^{2 \pi} dk_z G_{11}(k_x,k_y,k_z)\\
\end{eqnarray*}
where $\Lambda $ is momentum cutoff, and 
\newline
\newline
$G^{-1}$ =\[ \left( \begin{array}{lc}
\rho k^2 + \rho_z cos(k_z) + 2\lambda T &  -(2T \sigma +2A_0+ 2 \eta k_x k_y)\\
-(T \sigma +2A_0 \eta k_x k_y) &  \rho k^2 + \rho_z cos(k_z) + 2\lambda T
 \end{array} \right)\]
 
 %When $J_1<J_2$, $\eta$ has little effect on the solution.  Ignoring $\eta$, we can derive the following equations,
 To simplify the calculations\cite{note2}, we set $\eta=0$, in which case the integrals can be evaluated to yield
\begin{eqnarray*}
\frac{8\pi\rho\sigma}{g}&=&%\frac{1}{8\pi\rho}
 ln\left[\frac{\tilde\rho+\lambda-\sigma^\prime+\sqrt{\left(\tilde\rho+\lambda-\sigma^\prime\right)^2-\left(\tilde\rho_z\right)^2}}{
\lambda-\sigma^\prime+\sqrt{\left(\lambda-\sigma^\prime\right)^2-\left(\tilde\rho_z\right)^2}}\right]\\
&&-ln\left[\frac{\tilde\rho+\lambda+\sigma^\prime+\sqrt{(\tilde\rho+\lambda+\sigma^\prime)^2-\left(\tilde\rho_z\right)^2}}{
\lambda+\sigma^\prime+\sqrt{(\lambda+\sigma^\prime)^2-\left(\tilde\rho_z\right)^2}}\right],\\
\frac{8\pi\rho}{T}&=&ln\left[\frac{\tilde\rho+\lambda-\sigma^\prime+\sqrt{(\tilde\rho+\lambda-\sigma^\prime)^2-\left(\tilde\rho_z\right)^2}}{
\lambda-\sigma^\prime+\sqrt{(\lambda-\sigma^\prime)^2-\left(\tilde\rho_z\right)^2}}\right]\\
&&+ln\left[\frac{\tilde\rho+\lambda+\sigma^\prime+\sqrt{(\tilde\rho+\lambda+\sigma^\prime)^2-(\tilde\rho_z)^2}}{
\lambda+\sigma^\prime+\sqrt{(\lambda+\sigma^\prime)^2-\left(\tilde\rho_z\right)^2}}\right]
 \end{eqnarray*}
where the $\sigma^\prime=\sigma+A_0/T$, $\tilde \rho=\rho\Lambda^2/T$ and $\tilde \rho_z = \rho_z/T$. 

\begin{figure}[h!]
\centering
\caption{ Plot of the %N
nematic %O
order %P
parameter ($\sigma$) %with T
vs. temperature for different values of external field $A_0$ = 0 (red), 0.001 (green),0.01 (blue).  The parameters chosen are $\rho\Lambda^2=%\Lambda=
1,g=0.3$ and $\rho_z =0.01$. }
\includegraphics[width=0.5\textwidth]{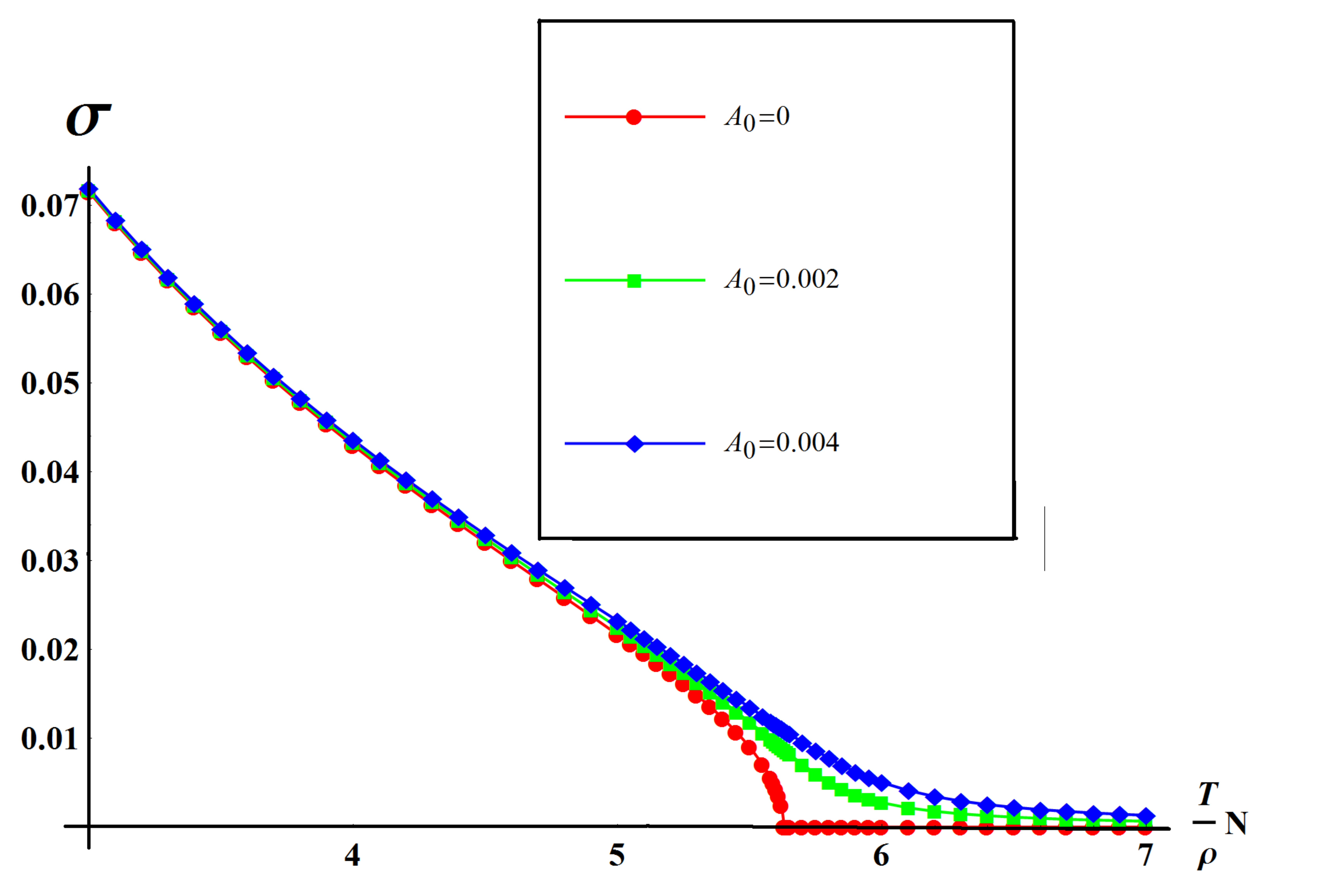}
\label{fig1}
\end{figure}

\begin{figure}[h!]
\centering
\caption{ Plot of the derivative of the nematic order parameter vs. temperature for different external fields $A_0(10^{-3})$. The parameters used are $\rho_z=0.01, g=0.3,\rho = \Lambda =1$. }
\includegraphics[width=0.5\textwidth]{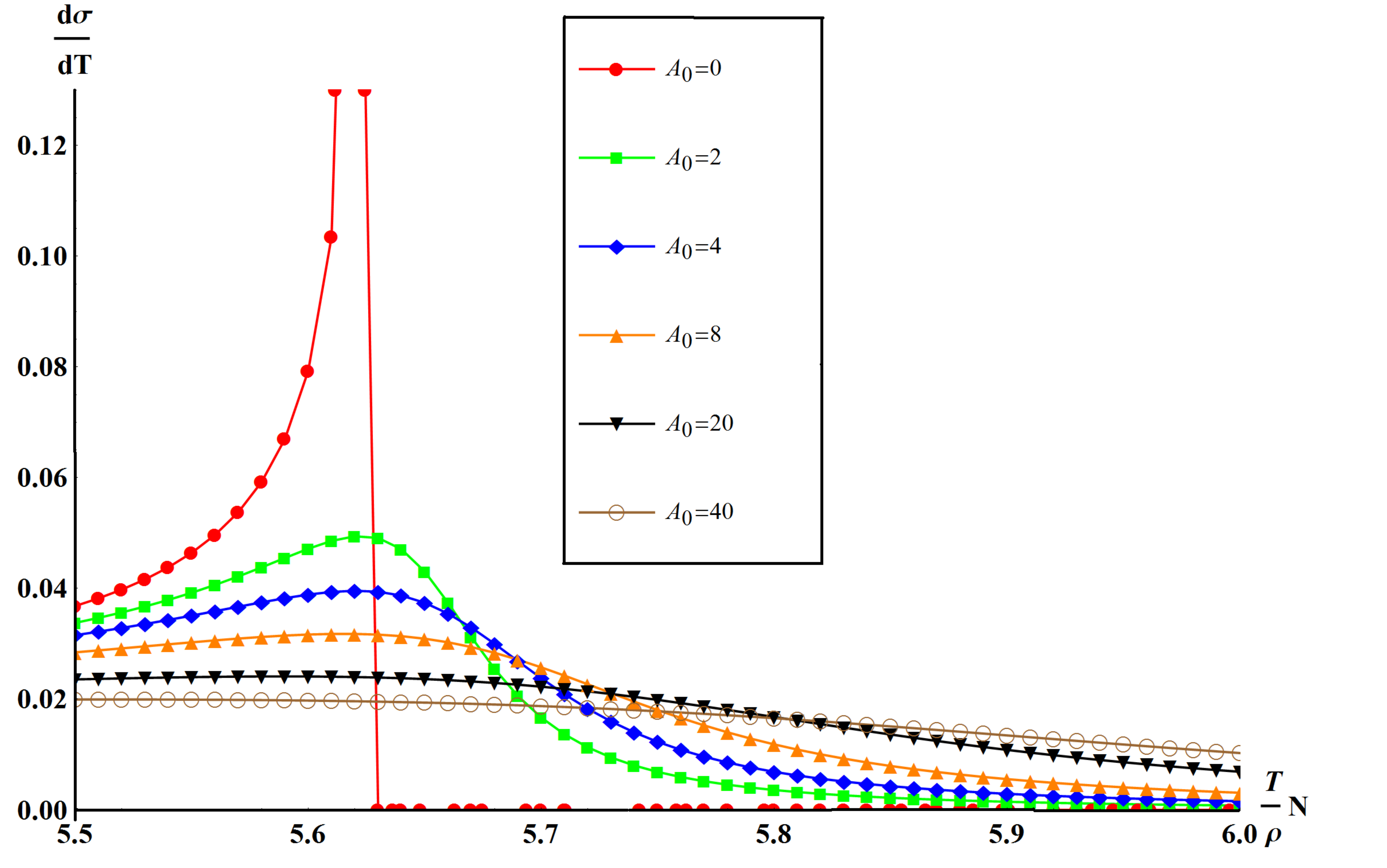}
\label{fig2}
\end{figure}

{\it Solutions:} In the absence of $A_0$, for any $g\neq 0$ there are two transition temperatures as shown in \cite{kivelson2008}. The nematic transition temperature $T_{\cal N}$ is determined by the discontinuity of the function $\frac{d\sigma}{dT}$ and the magnetic transition temperature $T_{AF}$ is determined by $\lambda= \sigma%_a
 +\tilde \rho_z$. However, when $A_0\neq0$, there is no discontinuity in the  function $\frac{d\sigma}{dT}$ because the external strain field already breaks the rotational symmetry\cite{note3}.  Typical %solutions of 
 plots $\sigma$  and $\frac{d\sigma}{dT}$ as a function of $T$ for different values of $A_0$ are shown in Fig.\ref{fig1} and Fig.\ref{fig2}. %respectively.  
 When $A_0$ is small,  a crossover temperature $T_{\cal N}^*$ can still be identified %in the curve of
 as the  inflection point temperature at which  $\frac{d\sigma}{dT}$ %where the derivative reaches its
 has a maximum. When $A_0$ is large, $\frac{d\sigma}{dT}$ increases smoothly as the temperature is lowered, so %. In this case, a 
 a well defined crossover temperature %can not be signed.  
 cannot be identified.  From the numerical data, 
we see that $A_0\sim 10^{-3}\rho$ is sufficiently large to %destroy 
%significantly
 %round the nematic crossover % behavior.   
% sufficiently that there is no
eliminate the inflection point.

Taking  $\lambda= \sigma_a+\rho_z/T$, we can derive the magnetic transition $T_{AF}$, which   is determined by the following equations,
\begin{eqnarray*}
\frac{\sigma_{AF}}{g}&=& \frac{1}{8\pi \rho}ln \left[\frac{2 \sigma_%{Na} 
{AF}^\prime T_{AF}+%\frac
{ \rho_z}%{T_{AF}} 
+ 2 \sqrt{\left(\sigma_{AF}^\prime T_{AF}\right)%{Na}
^2 + %\frac
{\sigma_%{Na}
{AF}^\prime \rho_z}%{T_{AF}}
}}{\rho_z%/(T_{AF})
} \right]\\
\frac{\sigma_{AF}}{g} &=& \frac{1}{4 \pi \rho}ln\frac{2 \rho}{\rho_z} - \frac{1}{T_{AF}}
\label{eqn3}
\end{eqnarray*}
where $\sigma_{AF}^\prime%{Na}
 = \sigma_{AF} + \frac{A_0}{T_{AF}}$.
%In the absence of the external stain field,    for iron-pnictides,  the parameters generally satisfy 
For $A_0=0$ and $\rho \Lambda^2 >g>>\rho_z$,  %The
the magnetic transition temperature $T^*_{AF}$ %can be approximated as
is approximately
\begin{eqnarray}
T^*_{AF}\approx T_{AF}^0+\frac{(T_{AF}^0)^2}{8\pi\rho}\left[ln\frac{4g}{\rho_z}+ln\left(\frac{T_{AF}^0}{8\pi\rho}ln\frac{4g}{\rho_z}\right)\right]
\end{eqnarray}
where $T_{AF}^0=\frac{4\pi \rho}{ln\frac{2\rho}{\rho_z}}$.
%In the presence of the 
For small external strain field, % it is reasonable to assume 
$A_0<<g$, % since a small $A_0$ can destroy the nematic crossover behavior as we have shown in fig.\ref{fig2}.
%In this case, we can define 
the shift in the magnetic ordering temperature, $ \Delta T_{AF} = T_{AF}(A_0)-T^*_{AF}$, %.  The leading order of $\Delta T_{AF}$ is given by
 to linear order is
\begin{eqnarray}
\Delta T_{AF}= \frac{T^*_{AF}}{(8\pi\rho-T^*_{AF})\sigma_{AF}-g}A_0.\label{eq}
\end{eqnarray}
% which suggests that the magnetic transition temperature increases linearly as increasing the external stain field. 
Since $\sigma_{AF}\sim g$, the coefficient in the right side of the above equation goes as $\sim 1/g$ for small $g$. %SAKsince $\sigma_{AF}\sim g$ as well. 
The above results can be extended even in the limit of   $g \to 0$.  It is easy to show that  in this limit,  $T_{\cal N}-T_{AF}\sim g^{1/2}$. Plugging this into Eq.\ref{eq}, we obtain  the expression below Eq. \ref{chi}.  

Eq. \ref{chi} can be further checked  in the case of $g=0$.  For $g=0$, %namely nematic transition is absent.  In this case, 
 the change of magnetic transition temperature is given by
\begin{eqnarray}
\Delta T_{AF}=\frac{J_z}{4\pi\rho(\frac{J_z}{T_{AF}^0})^{3/2}}\sqrt{T_{AF}^0|A_0|}
\end{eqnarray}
%which suggests the change of magnetic transition temperature is a square root function of the external strain field.
consistent with the expectations of scaling theory, Eq. \ref{chi}.

\textit{Comparison with experiment}:  Fig. 3 shows the comparison of our theoretical results with experimental observations of Dhital {\it et al} for the magnetic transition and nematic crossover as a function of uniaxial strain.   % Both the nematic and the %Neel T antiferromagnetic transition have the same trend as the pressure is coupled to Nematic order. 
The parameters used to generate the theoretical results, represented by the lines in the figure, are presented in the figure caption.  The upper curve, representing the nematic crossover, is solid where there is a well-defined inflection point associated with the crossover, and a dashed line where the crossover has become so smooth that there is no local maximum in the temperature derivative of the nematic order parameter.  The lower line indicates the theoretical magnetic ordering temperature.
 %The dashed line denotes the %expected trend in   the theoretical results as it is difficult to pinpoint exactly a nematic transition at large values of the external pressure field. 
 Fig. 4  shows the variation of the nematic transition with $A_0$ from which Fig. 3 was obtained.  The close correspondence between the theoretical and experimental curves supports the conjecture that the starting model captures the essential physics, although the comparison involves too many empirically determined parameters to make this conclusion inescapable. 
\begin{figure}[h!]
\centering
\caption{ Comparison of the external field, $A_0$, variation of Magnetic Ordering Temperature and Nematic transition temperature with experiment.The parameters chosen are $\rho_z=0.06 \rho, g =0.3 \rho, \rho \sim 71 K$. }
\includegraphics[width=0.5\textwidth]{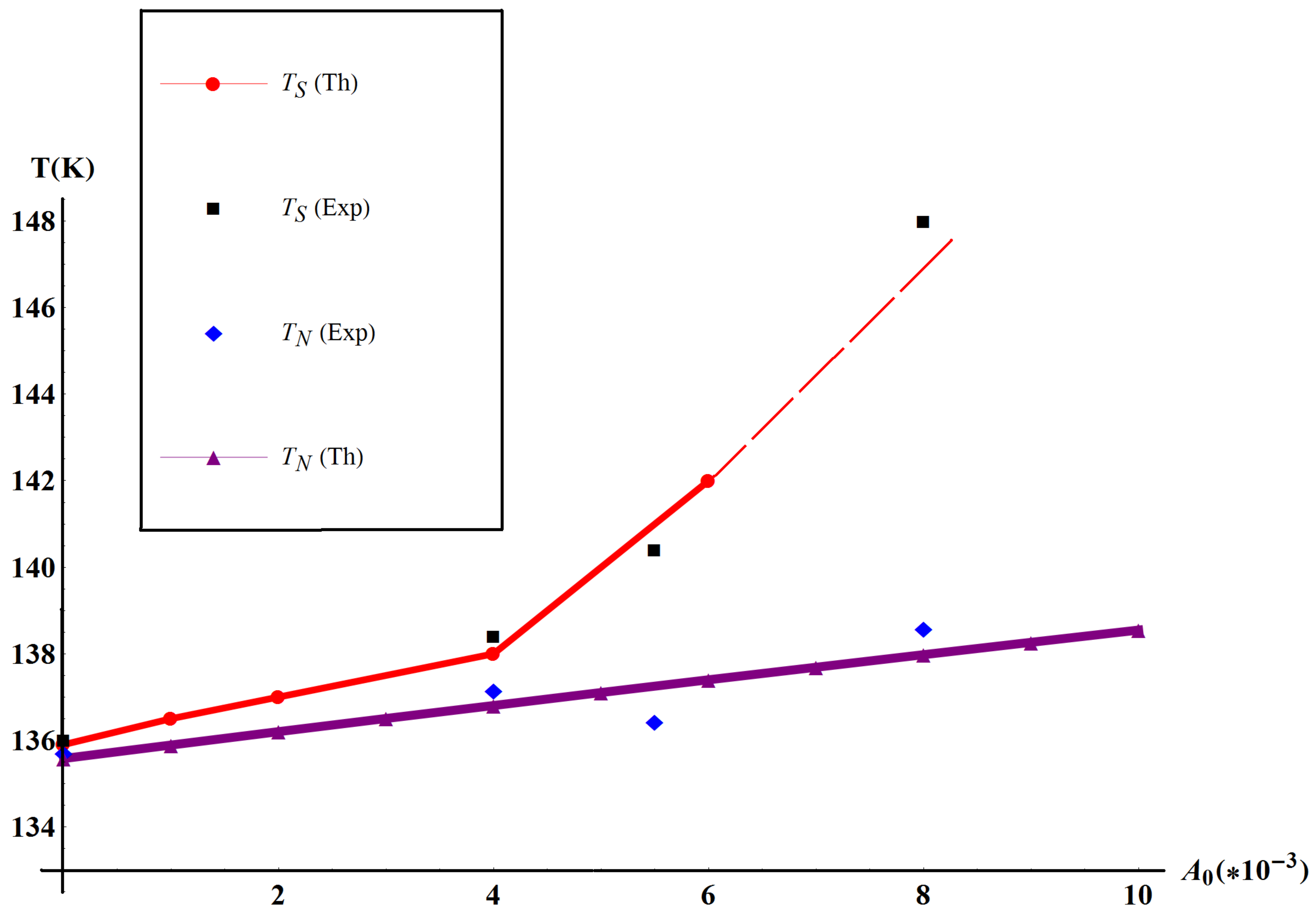}
\end{figure}
\begin{figure}[h!]
\centering
\caption{ Variation of the Nematic order jump with external field $A_0$. The parameters chosen are the same as Fig. 3 with N=3 and $A_0$ in units of $10^{-3}$.  }
\includegraphics[width=0.5\textwidth]{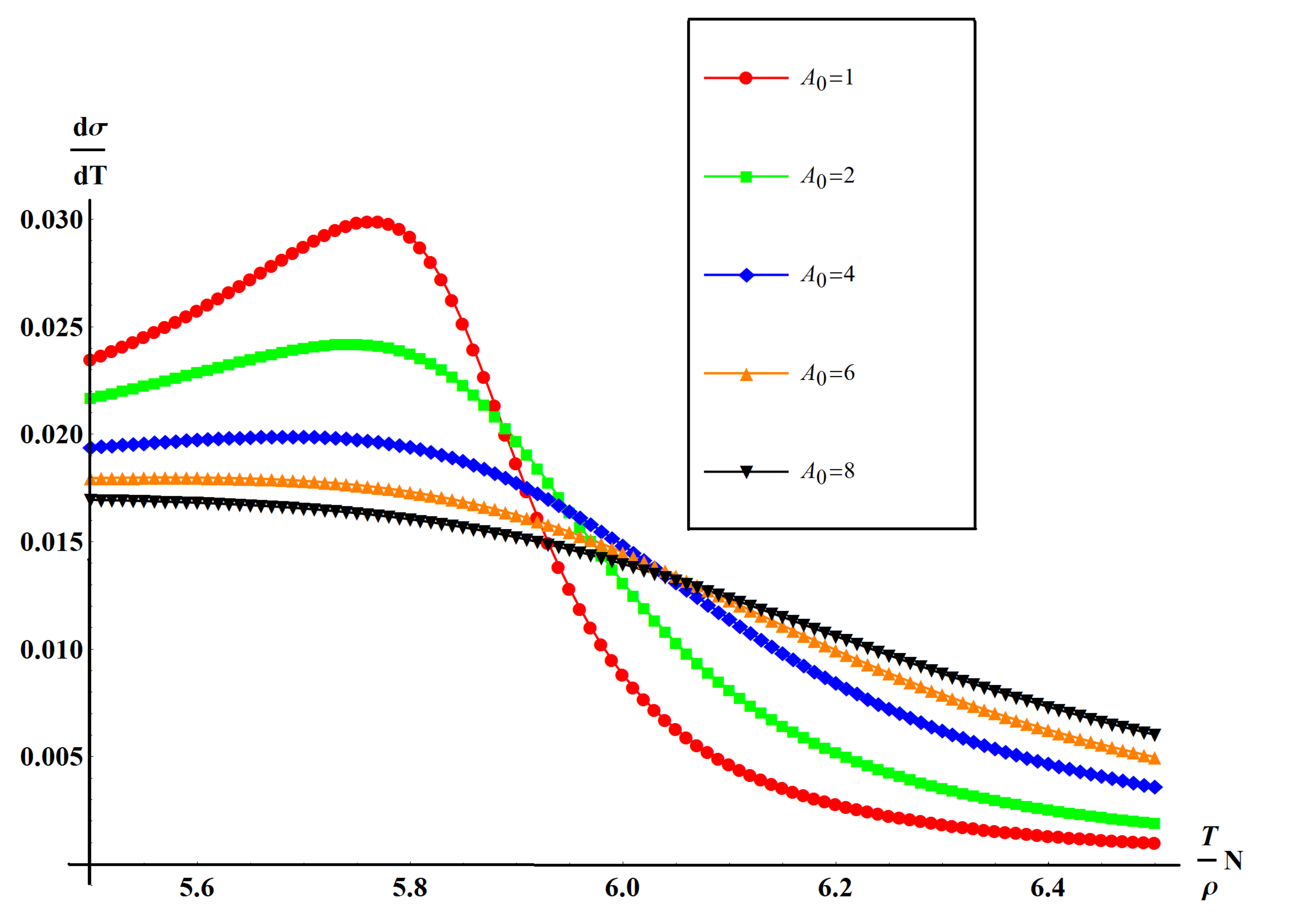}
\end{figure}
 In order to match the experimental transition temperature, the   value of $\rho$ used in our theoretical calculation ( see the caption of Fig. 3)  is smaller than the value measured in neutron scattering experiments\cite{Zhaojun2009}. This is expected since the transition temperature is always overestimated when $N$ is taken to be 3 in the large N %SAKlimit. 
 expression.

In summary, we %SAKshow
have shown that the minimal model with short ranged magnetic exchange couplings can   satisfactorily account for  the change of both structural and AF transition temperatures  under the uniaxial stain measured by Dhital {\it et al}  in  $BaFe_2As_2$. The experimental results strongly support the notion of  the magnetically-driven nematicity in iron-pnictides.

{\it Acknowledgement:}  The work is supported in part by the Ministry of Science and Technology of China 973 program(2012CV821400) and NSFC-1190024, and by DOE grant \# AC02-76SF00515 at Stanford (S. A. K.).


\begin{thebibliography}{10}
\expandafter\ifx\csname natexlab\endcsname\relax\def\natexlab#1{#1}\fi
\expandafter\ifx\csname bibnamefont\endcsname\relax
  \def\bibnamefont#1{#1}\fi
\expandafter\ifx\csname bibfnamefont\endcsname\relax
  \def\bibfnamefont#1{#1}\fi
\expandafter\ifx\csname citenamefont\endcsname\relax
  \def\citenamefont#1{#1}\fi
\expandafter\ifx\csname url\endcsname\relax
  \def\url#1{\texttt{#1}}\fi
\expandafter\ifx\csname urlprefix\endcsname\relax\def\urlprefix{URL }\fi
\providecommand{\bibinfo}[2]{#2}
\providecommand{\eprint}[2][]{\url{#2}}
 \bibitem{dhital}  Dhital C, Z. Yamani, Wei Tian, J. Zeretsky, A. S. Sefat, Ziqiang Wang, R. J. Birgeneau and  Stephen D. Wilson,  arXiv:1111.2326 (2011).
\bibitem[{\citenamefont{de~la Cruz et~al.}(2008)\citenamefont{de~la Cruz,
  Huang, Lynn, Li, II, Zarestky, Mook, Chen, Luo, Wang et~al.}}]{La1}
\bibinfo{author}{\bibfnamefont{C.}~\bibnamefont{de~la Cruz}},
  \bibinfo{author}{\bibfnamefont{Q.}~\bibnamefont{Huang}},
  \bibinfo{author}{\bibfnamefont{J.~W.} \bibnamefont{Lynn}},
  \bibinfo{author}{\bibfnamefont{J.}~\bibnamefont{Li}},
  \bibinfo{author}{\bibfnamefont{W.~R.} \bibnamefont{II}},
  \bibinfo{author}{\bibfnamefont{J.~L.} \bibnamefont{Zarestky}},
  \bibinfo{author}{\bibfnamefont{H.~A.} \bibnamefont{Mook}},
  \bibinfo{author}{\bibfnamefont{G.~F.} \bibnamefont{Chen}},
  \bibinfo{author}{\bibfnamefont{J.~L.} \bibnamefont{Luo}},
  \bibinfo{author}{\bibfnamefont{N.~L.} \bibnamefont{Wang}},
  \bibnamefont{et~al.}, \bibinfo{journal}{Nature}
  \textbf{\bibinfo{volume}{453}}, \bibinfo{pages}{899} (\bibinfo{year}{2008}).
\bibitem[{\citenamefont{Zhao et~al.}(2008{\natexlab{a}})\citenamefont{Zhao,
  Huang, de~la Cruz, Li, Lynn, Chen, Green, Chen, Li, Li et~al.}}]{Ce1}
\bibinfo{author}{\bibfnamefont{J.}~\bibnamefont{Zhao}},
  \bibinfo{author}{\bibfnamefont{Q.}~\bibnamefont{Huang}},
  \bibinfo{author}{\bibfnamefont{C.}~\bibnamefont{de~la Cruz}},
  \bibinfo{author}{\bibfnamefont{S.}~\bibnamefont{Li}},
  \bibinfo{author}{\bibfnamefont{J.~W.} \bibnamefont{Lynn}},
  \bibinfo{author}{\bibfnamefont{Y.}~\bibnamefont{Chen}},
  \bibinfo{author}{\bibfnamefont{M.~A.} \bibnamefont{Green}},
  \bibinfo{author}{\bibfnamefont{G.~F.} \bibnamefont{Chen}},
  \bibinfo{author}{\bibfnamefont{G.}~\bibnamefont{Li}},
  \bibinfo{author}{\bibfnamefont{Z.}~\bibnamefont{Li}}, \bibnamefont{et~al.},
  \bibinfo{journal}{Nature Materials} \textbf{\bibinfo{volume}{7}},
  \bibinfo{pages}{953} (\bibinfo{year}{2008}{\natexlab{a}}).

\bibitem{comment}  In fact, the symmetry involved is typically not a simple $\pi/2$ rotation, but rather a $\pi/2$ rotation followed by reflection through the Fe plane - this subtlety does not affect the following discussion.
\bibitem[{\citenamefont{Fang et~al.}(2008{\natexlab{a}})\citenamefont{Fang,
  Yao, Tsai, Hu, and Kivelson}}]{kivelson2008}
\bibinfo{author}{\bibfnamefont{C.}~\bibnamefont{Fang}},
  \bibinfo{author}{\bibfnamefont{H.}~\bibnamefont{Yao}},
  \bibinfo{author}{\bibfnamefont{W.-F.} \bibnamefont{Tsai}},
  \bibinfo{author}{\bibfnamefont{J.}~\bibnamefont{Hu}}, \bibnamefont{and}
  \bibinfo{author}{\bibfnamefont{S.~A.} \bibnamefont{Kivelson}},
  \bibinfo{journal}{Phys. Rev. B} \textbf{\bibinfo{volume}{77}},
  \bibinfo{pages}{224509} (\bibinfo{year}{2008}{\natexlab{a}}).

\bibitem[{\citenamefont{Xu et~al.}(2008)\citenamefont{Xu, Mueller, , and
  Sachdev}}]{xms2008}
\bibinfo{author}{\bibfnamefont{C.}~\bibnamefont{Xu}},
  \bibinfo{author}{\bibfnamefont{M.}~\bibnamefont{Mueller}}, 
  \bibnamefont{and} \bibinfo{author}{\bibfnamefont{S.}~\bibnamefont{Sachdev}},
  \bibinfo{journal}{Phys. Rev. B} \textbf{\bibinfo{volume}{78}},
  \bibinfo{pages}{20501} (\bibinfo{year}{2008}).
\bibitem{huxureview} Jiangping Hu and Cenke Xu,  arXiv:1112.2713 (2011).

\bibitem[{\citenamefont{Fradkin et~al.}(2010)\citenamefont{Fradkin, Kivelson,
  Lawler, Eisenstein, and Mackenzie}}]{fradkin-nematic}
\bibinfo{author}{\bibfnamefont{E.}~\bibnamefont{Fradkin}},
  \bibinfo{author}{\bibfnamefont{S.~A.} \bibnamefont{Kivelson}},
  \bibinfo{author}{\bibfnamefont{M.~J.} \bibnamefont{Lawler}},
  \bibinfo{author}{\bibfnamefont{J.~P.} \bibnamefont{Eisenstein}},
  \bibnamefont{and} \bibinfo{author}{\bibfnamefont{A.~P.}
  \bibnamefont{Mackenzie}}, \bibinfo{journal}{Annual Review of Condensed Matter
  Physics} \textbf{\bibinfo{volume}{1}}, \bibinfo{pages}{153}
  (\bibinfo{year}{2010}).
\bibitem[{\citenamefont{Fisher et~al.}(2011)\citenamefont{Fisher, Degiorgi, and
Shen}}]{nematic-fisher2011}
\bibinfo{author}{\bibfnamefont{I.~R.} \bibnamefont{Fisher}},
\bibinfo{author}{\bibfnamefont{L.}~\bibnamefont{Degiorgi}}, \bibnamefont{and}
\bibinfo{author}{\bibfnamefont{Z.~X.} \bibnamefont{Shen}},
\bibinfo{journal}{Reports on Progress in Physics}
\textbf{\bibinfo{volume}{74}}, \bibinfo{pages}{124506}
(\bibinfo{year}{2011}).
\bibitem[{\citenamefont{Ying et~al.}(2011)\citenamefont{Ying, Wang, Wu, Xiang,
Liu, Yan, Wang, Zhang, Ye, Cheng et~al.}}]{ying2011chen}
\bibinfo{author}{\bibfnamefont{J.~J.} \bibnamefont{Ying}},
\bibinfo{author}{\bibfnamefont{X.~F.} \bibnamefont{Wang}},
\bibinfo{author}{\bibfnamefont{T.}~\bibnamefont{Wu}},
\bibinfo{author}{\bibfnamefont{Z.~J.} \bibnamefont{Xiang}},
\bibinfo{author}{\bibfnamefont{R.~H.} \bibnamefont{Liu}},
\bibinfo{author}{\bibfnamefont{Y.~J.} \bibnamefont{Yan}},
\bibinfo{author}{\bibfnamefont{A.~F.} \bibnamefont{Wang}},
\bibinfo{author}{\bibfnamefont{M.}~\bibnamefont{Zhang}},
\bibinfo{author}{\bibfnamefont{G.~J.} \bibnamefont{Ye}},
\bibinfo{author}{\bibfnamefont{P.}~\bibnamefont{Cheng}},
\bibnamefont{et~al.}, \bibinfo{journal}{Physical Review Letters}
\textbf{\bibinfo{volume}{107}}, \bibinfo{pages}{067001}
(\bibinfo{year}{2011}).

\bibitem[{\citenamefont{Chuang et~al.}(2010)\citenamefont{Chuang, Allan, Lee,
Xie, Ni, Bud'ko, Boebinger, Canfield, and Davis}}]{nematic-chuang2010}
\bibinfo{author}{\bibfnamefont{T.-M.} \bibnamefont{Chuang}},
\bibinfo{author}{\bibfnamefont{M.~P.} \bibnamefont{Allan}},
\bibinfo{author}{\bibfnamefont{J.}~\bibnamefont{Lee}},
\bibinfo{author}{\bibfnamefont{Y.}~\bibnamefont{Xie}},
\bibinfo{author}{\bibfnamefont{N.}~\bibnamefont{Ni}},
\bibinfo{author}{\bibfnamefont{S.~L.} \bibnamefont{Bud'ko}},
\bibinfo{author}{\bibfnamefont{G.~S.} \bibnamefont{Boebinger}},
\bibinfo{author}{\bibfnamefont{P.~C.} \bibnamefont{Canfield}},
\bibnamefont{and} \bibinfo{author}{\bibfnamefont{J.~C.} \bibnamefont{Davis}},
\bibinfo{journal}{Science} \textbf{\bibinfo{volume}{327}},
\bibinfo{pages}{181} (\bibinfo{year}{2010}).
\bibitem[{\citenamefont{Zhao et~al.}(2009)\citenamefont{Zhao, Adroja, Yao,
Bewley, Li, Wang, Wu, Chen, Hu, and Dai}}]{Zhaojun2009}
\bibinfo{author}{\bibfnamefont{J.}~\bibnamefont{Zhao}},
\bibinfo{author}{\bibfnamefont{D.~T.} \bibnamefont{Adroja}},
\bibinfo{author}{\bibfnamefont{D.~X.} \bibnamefont{Yao}},
\bibinfo{author}{\bibfnamefont{R.}~\bibnamefont{Bewley}},
\bibinfo{author}{\bibfnamefont{S.~L.} \bibnamefont{Li}},
\bibinfo{author}{\bibfnamefont{X.~F.} \bibnamefont{Wang}},
\bibinfo{author}{\bibfnamefont{G.}~\bibnamefont{Wu}},
\bibinfo{author}{\bibfnamefont{X.~H.} \bibnamefont{Chen}},
\bibinfo{author}{\bibfnamefont{J.~P.} \bibnamefont{Hu}}, \bibnamefont{and}
\bibinfo{author}{\bibfnamefont{P.~C.} \bibnamefont{Dai}},
\bibinfo{journal}{Nature Physics} \textbf{\bibinfo{volume}{5}},
\bibinfo{pages}{555} (\bibinfo{year}{2009}).
\bibitem[{\citenamefont{Harriger et~al.}(2010)\citenamefont{Harriger, Luo, Liu,
Perring, Frost, Hu, Norman, and Dai}}]{Harringer2010}
\bibinfo{author}{\bibfnamefont{L.~W.} \bibnamefont{Harriger}},
\bibinfo{author}{\bibfnamefont{H.}~\bibnamefont{Luo}},
\bibinfo{author}{\bibfnamefont{M.}~\bibnamefont{Liu}},
\bibinfo{author}{\bibfnamefont{T.~G.} \bibnamefont{Perring}},
\bibinfo{author}{\bibfnamefont{C.}~\bibnamefont{Frost}},
\bibinfo{author}{\bibfnamefont{J.}~\bibnamefont{Hu}},
\bibinfo{author}{\bibfnamefont{M.~R.} \bibnamefont{Norman}},
\bibnamefont{and} \bibinfo{author}{\bibfnamefont{P.}~\bibnamefont{Dai}},
\bibinfo{journal}{Phys. Rev B} \textbf{\bibinfo{volume}{84}},
\bibinfo{pages}{054544} (\bibinfo{year}{2010}).
\bibitem[{\citenamefont{Nakajima et~al.}(2011)\citenamefont{Nakajima, Liang,
Ishida, Tomioka, Kihou, Lee, Iyo, Eisaki, Kakeshita, Ito
et~al.}}]{Nakajima-nematic-optical}
\bibinfo{author}{\bibfnamefont{M.}~\bibnamefont{Nakajima}},
\bibinfo{author}{\bibfnamefont{T.}~\bibnamefont{Liang}},
\bibinfo{author}{\bibfnamefont{S.}~\bibnamefont{Ishida}},
\bibinfo{author}{\bibfnamefont{Y.}~\bibnamefont{Tomioka}},
\bibinfo{author}{\bibfnamefont{K.}~\bibnamefont{Kihou}},
\bibinfo{author}{\bibfnamefont{C.~H.} \bibnamefont{Lee}},
\bibinfo{author}{\bibfnamefont{A.}~\bibnamefont{Iyo}},
\bibinfo{author}{\bibfnamefont{H.}~\bibnamefont{Eisaki}},
\bibinfo{author}{\bibfnamefont{T.}~\bibnamefont{Kakeshita}},
\bibinfo{author}{\bibfnamefont{T.}~\bibnamefont{Ito}}, \bibnamefont{et~al.},
\bibinfo{journal}{PNAS} \textbf{\bibinfo{volume}{108}},
\bibinfo{pages}{12238} (\bibinfo{year}{2011}).
\bibitem[{\citenamefont{Yi et~al.}(2011{\natexlab{a}})\citenamefont{Yi, Lu,
Chu, Analytis, Sorini, Kemper, Mo, Moore, Hashimoto, Lee
et~al.}}]{nematic-yi2010}
\bibinfo{author}{\bibfnamefont{M.}~\bibnamefont{Yi}},
\bibinfo{author}{\bibfnamefont{D.~H.} \bibnamefont{Lu}},
\bibinfo{author}{\bibfnamefont{J.-H.} \bibnamefont{Chu}},
\bibinfo{author}{\bibfnamefont{J.~G.} \bibnamefont{Analytis}},
\bibinfo{author}{\bibfnamefont{A.~P.} \bibnamefont{Sorini}},
\bibinfo{author}{\bibfnamefont{A.~F.} \bibnamefont{Kemper}},
\bibinfo{author}{\bibfnamefont{S.-K.} \bibnamefont{Mo}},
\bibinfo{author}{\bibfnamefont{R.~G.} \bibnamefont{Moore}},
\bibinfo{author}{\bibfnamefont{M.}~\bibnamefont{Hashimoto}},
\bibinfo{author}{\bibfnamefont{W.~S.} \bibnamefont{Lee}},
\bibnamefont{et~al.}, \bibinfo{journal}{ArXiv:1011.0050}
(\bibinfo{year}{2011}{\natexlab{a}}).
\bibitem[{\citenamefont{Yi et~al.}(2011{\natexlab{b}})\citenamefont{Yi, Lu,
Moore, Kihou, Lee, Iyo, Eisaki, Yoshida, Fujimori, and
Shen}}]{Yi2011-nematic}
\bibinfo{author}{\bibfnamefont{M.}~\bibnamefont{Yi}},
\bibinfo{author}{\bibfnamefont{D.~H.} \bibnamefont{Lu}},
\bibinfo{author}{\bibfnamefont{R.~G.} \bibnamefont{Moore}},
\bibinfo{author}{\bibfnamefont{K.}~\bibnamefont{Kihou}},
\bibinfo{author}{\bibfnamefont{C.-H.} \bibnamefont{Lee}},
\bibinfo{author}{\bibfnamefont{A.}~\bibnamefont{Iyo}},
\bibinfo{author}{\bibfnamefont{H.}~\bibnamefont{Eisaki}},
\bibinfo{author}{\bibfnamefont{T.}~\bibnamefont{Yoshida}},
\bibinfo{author}{\bibfnamefont{A.}~\bibnamefont{Fujimori}}, \bibnamefont{and}
\bibinfo{author}{\bibfnamefont{Z.-X.} \bibnamefont{Shen}},
\bibinfo{journal}{Arxiv:1111.6134} (\bibinfo{year}{2011}{\natexlab{b}}).

\bibitem{fernandez}   R. M. Fernandes, E. Abrahams, and J. Schmalian, Phys. Rev. Lett. {\bf 107}, 217002 (2011) and R.M. Fernandes, A. V. Chubukov, J. Knolle, I. Ermin, and J. Schmalian, arXiv:1110.1893 (2011).

\bibitem[{\citenamefont{Wysocki et~al.}(2011)\citenamefont{Wysocki,
  Belashchenko, and Antropov}}]{Wysocki2011}
\bibinfo{author}{\bibfnamefont{A.~L.} \bibnamefont{Wysocki}},
  \bibinfo{author}{\bibfnamefont{K.~D.} \bibnamefont{Belashchenko}},
  \bibnamefont{and} \bibinfo{author}{\bibfnamefont{V.~P.}
  \bibnamefont{Antropov}}, \bibinfo{journal}{Nature Physics}
  \textbf{\bibinfo{volume}{7}}, \bibinfo{pages}{485} (\bibinfo{year}{2011}).
\bibitem[{\citenamefont{Hu et~al.}(2011)\citenamefont{Hu, Xu, Liu, Hao, and
  Wang}}]{magnetic-hu2011}
\bibinfo{author}{\bibfnamefont{J.}~\bibnamefont{Hu}},
  \bibinfo{author}{\bibfnamefont{B.}~\bibnamefont{Xu}},
  \bibinfo{author}{\bibfnamefont{W.}~\bibnamefont{Liu}},
  \bibinfo{author}{\bibfnamefont{N.}~\bibnamefont{Hao}}, \bibnamefont{and}
  \bibinfo{author}{\bibfnamefont{Y.}~\bibnamefont{Wang}},
  \bibinfo{journal}{ArXiv:1106.5169}  (\bibinfo{year}{2011}).
\bibitem{note2} So long as $\eta$ is not too large, the large $N$ approach can be straightforwardly extended to include the effects of non-zero $\eta$, but it has little effect on the results.
\bibitem{note3} Under some circumstances, there could be a metanematic transition, even for $A_0> 0$, but this does not seem to occur in the $N\to\infty$ limit.






















\end{thebibliography}
\end{document}